\documentclass[useAMS,usenatbib]{mn2e}

\usepackage{txfonts}

\usepackage{graphicx}
\usepackage{xcolor}
\usepackage{epsfig,graphics}
\usepackage{rotating} 
\usepackage{epsfig}

\usepackage{graphicx,natbib,multirow,topcapt}
\usepackage{savefnmark}
    
\newcommand {\be}{\begin{equation}}
\newcommand {\ee}{\end{equation}}  
\renewcommand{\theenumi}{\arabic{enumi}.}

\newcommand{\doz}{\left( z \right)}
\newcommand{\oaz}{\left(1+ z \right)}
\newcommand{\dozsub}[1]{\left( z_{ #1 } \right)}
\newcommand{\oazsub}[1]{\left(1+ z_{ #1 } \right)}
\newcommand{\e}{eROSITA}
\newcommand{\ledd}{L_{Edd}}
\newcommand{\edd}{_{Edd}}

\begin{document}

\title[\textit{SRG}/eROSITA prospects for TDE
  detection]{\textit{SRG}/eROSITA prospects for the detection of stellar
  tidal disruption flares}  
\author[I. Khabibullin et al.]{I. Khabibullin$^{1,2}$\thanks{E-mail:
khabibullin@iki.rssi.ru}, S. Sazonov$^{1,3,2}$ and R. Sunyaev$^{2,1}$\\
$^{1}$Space Research Institute, Russian Academy of Sciences,
Profsoyuznaya 84/32, 117997 Moscow, Russia\\
$^{2}$Max-Planck-Institut f\"ur Astrophysik,
Karl-Schwarzschild-Str. 1, 85740 Garching bei M\"unchen, Germany\\
$^{3}$Moscow Institute of Physics and Technology, Institutsky
per. 9, 141700 Dolgoprudny, Russia
}

\maketitle

\begin{abstract}

We discuss the potential of the \e\ telescope on board the
\textit{Spectrum-Roentgen-Gamma (SRG)} observatory to detect stellar tidal
disruption events (TDE) during its 4-year all-sky survey.  These
  events are expected to reveal themselves as luminous flares of
  UV/soft X-ray emission associated with the centers of previously
  non-active galaxies and fading by few orders of magnitude
  on time-scales of several months to years. Given that
  \e\ will complete an all-sky survey every 6~months and a total of
  8~such scans will be performed over the course of the mission,
we propose
to distinguish TDEs from other X-ray transients using two
criteria: i) large (more than a factor of 10) X-ray variation
between two subsequent 6-month scans and ii) soft X-ray
spectrum. The expected number of TDE candidates is $\sim 10^3$ per
scan (with most of the events being new discoveries in a given scan),
so that a total of several thousand TDE candidates could be found during 
the 4-year survey. The actual number may significantly differ from
this estimate, since it is based on just a few TDEs observed so
far. The \e\ all-sky survey is expected to be nearly equally
sensitive to TDEs occurring near supermassive black holes (SMBH) of
mass between $\sim 10^6$ and $\sim 10^7M_\odot$ and will thus provide
a unique census of quiescent SMBHs and associated nuclear stellar
cusps in the local Universe ($z\lesssim 0.15$). Information on TDE
candidates will be available within a day after their detection and
localization by \e, making possible follow-up observations that
may reveal peculiar types of TDEs.

\end{abstract}

\section{Introduction}
\label{s:intro}

It is widely accepted that supermassive black holes (SMBH),
i.e. black holes with mass of hundreds of thousands to billions of
solar masses, reside in the centers of most galaxies (see
\citealt{Ho2008} for a review). However, apart from active galactic
nuclei (AGN), where intense accretion of gas onto the SMBH takes
place,  and the nuclei of the few most nearby optically non-active galaxies, where
dynamics of stars and gas near the SMBH can be directly measured, such
black holes remain hidden \citep{Kormendy1995,Kormendy2001}. A
plausible scenario for revealing a compact object with 
$M\lesssim 10^8 M_{\odot} $\footnote{SMBHs with $M\gtrsim 10^8
  M_{\odot}$ are expected to capture solar-type stars 
without disruption, since the corresponding tidal radius is smaller
than the radius of the black hole's event horizon. Also, in the
situation where the tidal and horizon radii are close to each other,
general relativity effects cause the capture probability to
strongly depend on the black hole spin \citep{Kesden2012}.} in the center 
of an inactive galaxy is based on tidal disruption of stars passing
sufficiently close to it \citep{Hills1975,Lidskii1979}. According to
analytical estimates, about half of the disrupted star's material
should be captured and then accreted by the SMBH on a time-scale of
about a year \citep{Gurzadian1981,Rees1988}. Numerical simulations
confirm these suggestions \citep{Evans1989,Laguna1993}, though
  \cite{Ayal2000} found that only about 10\% of the original star's
  mass actually gets accreted. Accretion of the stellar debris onto a
  SMBH gives rise to a flare of thermal radiation with the 
peak at extreme ultraviolet (EUV)/soft X-ray wavelengths 
and maximum luminosity of $10^{43}-10^{45}$ erg s$^{ -1} $\citep{Strubbe2009}. 
However, the expected rate of such tidal disruption events (TDE) is
not high: depending on the stellar density in the nuclear cusp and the
SMBH mass, it varies from $ 10^{-6}$ to $  10^{-3} $ yr$ ^{-1} $ per galaxy
\citep{Wang2004}. Consistent with these expectations, just
about a dozen of TDE candidates have been identified so far by X-ray
\citep{Komossa2002,Donley2002,Esquej2008,Cappelluti2009,Maksym2010,Lin2011,Saxton2012},  
UV  (\citealt{Gezari2009} and references therein) and optical 
(\citealt{vanVelzen2011A} and references therein) observations. The
observed spectral and temporal properties of these TDE candidates
  support the general picture of captured stellar debris forming an 
  accretion disk around a SMBH, but the existing data are too
sparse to provide stringent constraints on the theoretical models.  

Recently, it has also been recognized that powerful jets could emerge
during accretion of stellar debris onto a SMBH, giving rise to radio
and hard X-ray emission
\citep{Giannios2011,vanVelzen2011B,Lei2011,Krolik2012}. The subsequent
discovery by \textit{Swift} of TDE candidates that fit well into this 
scenario  \citep{Levan2011,Burrows2011,Cenko2012} opened an extra
dimension in TDE studies. It is currently unclear whether the observed
rarity of such flares is owing to a high degree of jet collimation or
to a low intrinsic probability of jet production
\citep{Cenko2012}. Obviously, a larger sample of TDEs is needed to
check the different possibilities and test the existing theoretical
models.    

Deep wide-area surveys are especially powerful in discovering
transient sources, usually by comparison with pre- or post-survey
observations. As regards giant thermal flares from stellar TDEs,
EUV/soft X-ray surveys are the most suitable. Up to several thousand
TDEs had been expected to be detected during the \textit{ROSAT} All-Sky
Survey (RASS, \citealt{Sembay1993}), but only \textit{five}
  were actually identified due to the lack of timely follow-up
  observations and deep X-ray observations to confirm the pre- or
  post-outburst state \citep{Komossa2002,Donley2002}. Nonetheless, the
  RASS has provided the largest contribution to the existing sample
  of X-ray TDE detections. The several other candidates have been
  found in the \textit{XMM-Newton} Slew Survey
  \citep{Esquej2008,Lin2011,Saxton2012} and in \textit{Chandra}
  pointed observations \citep{Cappelluti2009,Maksym2010}.

\begin{figure}
\centering
\includegraphics[width=\columnwidth]{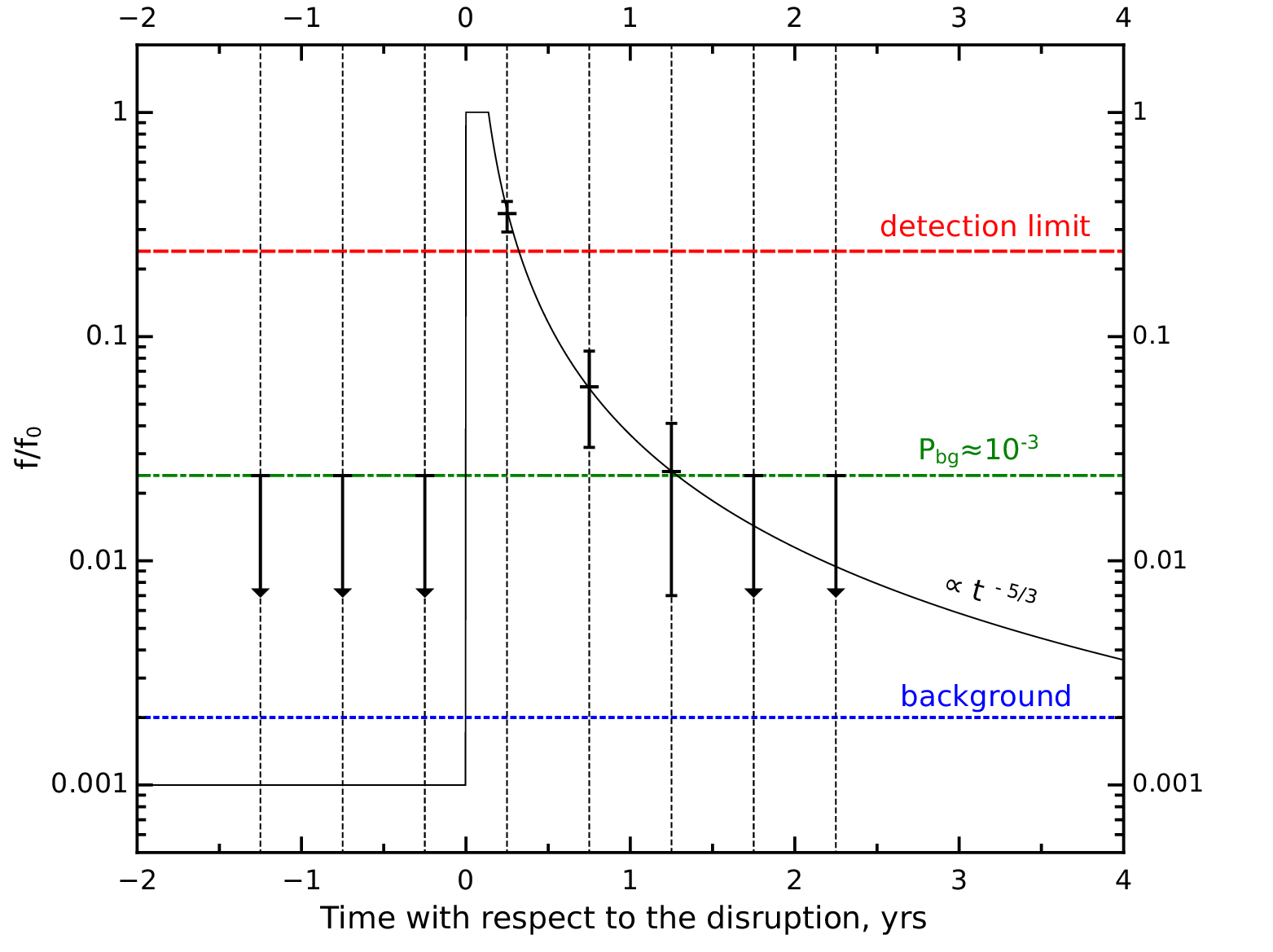}
\topcaption{Light curve of a TDE ~ candidate (normalized to the
    flux value at peak $ f_0 $) as would be seen by eROSITA. The 
  source falls into the telescope's field of view every 6~months, as
  indicated by the vertical lines. The average background level is
  shown by the blue dotted line. The green dash-dotted line marks
  the count rate that can be produced by Poisson fluctuations in the
  background with a probability $P_{bg}\simeq 10^{-3}$. The flux from
  a TDE candidate must exceed this level by a factor $\ge 10$ during
  at least one scan, which defines the TDE detection level shown
  by the red dashed line. See Appendix for further
  details. 
} 
\label{fig:obs}
\end{figure}

In the present paper, we investigate opportunities provided 
by the upcoming \textit{Spectrum-X-Gamma (SRG)} mission. This
observatory is planned to be launched in 2014 to the L2 point of the
Earth--Sun system and the first 4~years of the mission will be devoted
to an all-sky survey in the 0.2--12~keV energy band with the
\e\footnote{Extended ROentgen Survey with An Imaging Telescope Array} 
\citep{Merloni2012} and ART-XC\footnote{Astronomical Roentgen
  Telescope -- X-ray Concentrator} \citep{Pavlinsky2012}
telescopes. The survey will consist of 8 repeated half-a-year scans of
the entire sky, each with sensitivity (below 2~keV) $\sim 4$
  times better than RASS. Owing to this observational strategy,
'self-follow-up' 
observations of X-ray sources will be possible through scan-to-scan
comparison. As one of the main TDE hallmarks is a giant amplitude of
flux variations on the time-scale of years, searches for such events
in the \e\ All-Sky Survey (eRASS) data can be very efficient (see
Fig.~\ref{fig:obs}). An additional identification tool will be
provided by X-ray spectral analysis, taking advantage of the high
sensitivity in the 0.2--2~keV energy range and the possibility of
detection of harder photons with energies up to $\sim 10$~keV. Below,
after a short overview of TDE X-ray emission properties, we specify
the required identification criteria and estimate the expected TDE
detection rate for eRASS. 

\section{X-ray signatures of tidal disruption events} 
\label{s:sign}

The observational appearance of a given TDE may depend on the
properties of the star (mass and type) and SMBH (mass and spin) as
well as on the initial stellar orbit characteristics. To simplify the 
treatment, we shall use some `average' pattern which is based
  on the standard theory of tidal disruption of a \textit{Sun-like}
  star and which satisfactorily reflects major observational
  properties of the currently available sample of X-ray selected
  TDEs. Hence, such phenomena as flares resulting from tidal stripping of 
  atmospheres of giant stars \citep{MacLeod2012,Guillochon2013} or
  stellar tidal disruption flares from recoiling SMBHs
  \citep{Komossa2012} are not considered due to the lack of
  observational data associated with them (see \citealt{Alexander2012}
  for a review of various dynamical channels by which stars can be
  supplied to a SMBH).    


\subsection{Temporal properties}
\label{ss:temporal}

The flux decay of TDE flares is broadly consistent with a power law
with a slope of $\sim -5/3$
\citep{Komossa2002,Halpern2004,Vaughan2004}. This decay probably
reflects the decreasing accretion rate, $\dot{M}$, of the falling back
stellar material by the SMBH, coupled with some constant radiative
efficiency $\epsilon$ (the source luminosity $L=\epsilon \dot{M}c^2$),
since the viscous timescale is believed to be shorter than
the characteristic dynamical fallback time
\citep{Rees1988,Phinney1989}.  However, \cite{Lodato2009} 
  found that a $ t^{-5/3} $ decrease of $ \dot{M} $ is reached only
  asymptotically at late times, while at earlier stages the behavior
  of the accretion rate is sensitive to the structure of the disrupted
  star. Besides that, 
  the radiative efficiency of accretion is probably constant  
only at a certain phase of the event, namely when $\dot{M}$ is below
the critical value $\dot{M}_{Edd}=\epsilon^{-1} \ledd/c^2$ 
corresponding to the Eddington luminosity $\ledd \backsimeq 1.3 \times
10^{44}\left(\frac{M_{BH}}{10^6 M_{\sun}}\right)$~erg~s$^{-1}$, but still
above $\sim 0.01 \dot{M}_{Edd}$. At this stage, the released gravitational
energy is radiated away by a geometrically thin, optically thick
accretion disk \citep{SS1973}. For a solar type star with a
peri-center distance $R_p =3 R_{S}$ (where $R_S$ is the Schwarzschild
radius), the peak mass rate occurs at time 
\begin{equation}
\label{eq:ti}
\tau_{i}\backsimeq 20\left(\frac{M_{BH}}{10^6
  M_{\sun}}\right)^{5/2}~{\rm min}
\end{equation}
after the instant of disruption $t_0$ and is highly supercritical 
given $\epsilon=0.1$ and $M_{BH} \lesssim 10^7 M_{\sun}$
(\citealt{Strubbe2009}; this delay is in fact highly sensitive to
pericenter distance: $\tau_{i}\propto R_p^3$). Therefore, early on
accretion takes place in radiatively inefficient regime, with an 
approximately constant luminosity $ L\backsimeq \ledd $ emitted by a
thick accretion disk, which is likely subject to outflows
\citep{Ulmer1999,Strubbe2009}. The boundary between the early
(Eddington) and late (decay) phases lies at 
\begin{equation}
\label{eq:tedd}
\tau\edd\backsimeq 0.1 \left(\frac{M_{BH}}{10^6M_{\sun}}\right)^{2/5}~{\rm yr}
\end{equation}
after disruption for $R_p =3 R_{S}$ ($\tau\edd \propto R_p^{6/5}$,
\citealt{Strubbe2009}). Although the 
reality can be somewhat more complicated (see  
e.g. \citealt{Lodato2009}), we adopt the following approximate
relation for the time profile of flares associated with TDEs: 
\begin{equation}
\label{eq:decay}
	L(t)= 	\left\{
\begin{array}{ll}
L_{quies} & {\rm for}~ \ t<t_0 \\
L_0 & {\rm for}~ t_0<t<t_1 \\
L_0 \left(\frac{t-t_0}{\tau\edd}\right)^{-5/3}& {\rm for}~  t>t_1 
\end{array}
\right.
\end{equation}
where $L_{quies}$ is the source luminosity in the quiescent state,
$L_{0}=\eta\ledd\gg L_{quies}$ is the peak observed
  luminosity (with $\eta$ being a geometrical  
dilution factor), and $t_1=t_0+\tau\edd$. Equation (\ref{eq:decay})
ignores the short period $\tau_{i}\ll\tau\edd$ since
$\tau_{i}/{\tau\edd}\simeq 4\times 10^{-4}\left({M_{BH}}/{10^6 
M_{\sun}}\right)^{21/10}\left({R_{p}}/{3R_S}\right)^{9/5}$. However,
as we further discuss in Section~\ref{ss:depend} below, there is an
interesting possibility to observe some TDEs during the rising phase
of the flare, which lasts $\tau_{rise}<\tau_{i}\ll \tau\edd$ (under
our assumption of shortness of the viscous time compared to the
fallback time) and can be significantly long for large $M_{BH}$ and/or
$R_p$.

For SMBHs with $M_{BH}\gtrsim {\rm few} \times 10^7M_{\sun}$, only the
closest stellar disruptions (small $R_p$) can
lead to supercritical accretion, while the majority of TDE flares
associated with such SMBHs are expected to be sub-Eddington (given a
uniform distribution of pericenters, \citealt{Ulmer1999}). Moreover,
most of the emission will appear at UV rather than X-ray energies (see 
Section~\ref{ss:spectral} below). We thus restrict our consideration
to SMBHs with $M_{BH}\lesssim 10^7 M_{\sun}$. 

We note that very recent analytic \citep{Stone2012} and
  numerical \citep{Guillochon2013} calculations indicate that the
  characteristic timescales discussed above might be almost
  independent of $ R_p $, since the stellar debris energy 'freezes in'
  at the tidal radius $R_{t}=R_{\star} \left(M_{BH}/M_{\star}\right)^{1/3}$
  (where $R_{\star} $ and $M_{\star} $ are the mass and radius of the
  disrupted star), rather than at $ R_p $ as was thought before (see
  \citealt{Stone2012} for a thorough discussion). The resulting
  difference is small for $M_{BH}=10^7 M_{\sun}$ (since $R_t/3
  R_S=1.6$, assuming $R_{\star}=R_{\sun}$ and $M_{\star}=M_{\sun}$),
  but becomes significant for lighter SMBHs, since $R_t/3 R_S=7.5$ for  
  $M_{BH}\sim 10^6 M_{\sun}$, which should lead to longer $ \tau_{i} $ and
  $ \tau_{Edd} $ and hence to slower flux decay during the post-peak
  phase. However, there is no observational confirmation of 
  these new theoretical results yet. Furthermore, the TDE light curves
  observed so far are consistent with the shorter decay time-scales
  expected for $R_p\sim 3 R_S$ \citep{Esquej2008}. Since we make our
  predictions for the \e\ TDE detection rate (in Section 
  \ref{s:pred}) using the same (currently available) data
  set, we shall for simplicity assume $R_p=3 R_S$.

\subsection{Spectral properties}
\label{ss:spectral}

An observer will detect only a fraction of the TDE bolometric flux,
depending on the spectral energy distribution of TDE emission and
the energy response of the detector used. Since this emission
presumably originates in an accretion disk around a SMBH, its spectral 
energy distribution can be approximated \citep{SS1973} by a black body
radiation spectrum with temperature   
\be
\kappa T_{bb} \simeq 0.06 \left(\frac{M_{BH}}{10^6
  M_{\odot}}\right)^{-1/4} \left(\frac{\dot{M}}{\dot{M\edd}}
\right)^{1/4}\,{\rm keV}, 
\label{eq:temperature}
\ee
which corresponds to the innermost regions of the disk
($R \sim 5 R_{S}$, \citealt{Ulmer1999}). Since this emission is
fairly soft, the detection probability must strongly depend on the
absorption column density to the source, implying that X-ray
selected samples might be biased against absorbed TDE events. No
indication of significant intrinsic absorption has been found from 
spectral analysis performed for known TDE candidates under the
assumption of black body emission. The estimated black body
temperatures are $ \kappa T_{bb} \simeq 0.07$ keV (the median value,
\citealt{Komossa2002, Esquej2008}), although the quality of the
existing X-ray data hardly makes it possible to distinguish a black body
model from e.g. an absorbed power-law model
\citep{Esquej2008}. Power-law fits typically resulted in steep slopes
($\Gamma \gtrsim 3$), i.e. the observed spectra are indeed soft.

Disregarding the remaining uncertainty in the spectral shape of TDE
emission, we use the (physically motivated) absorbed black body model
in our calculations. This model has 3 parameters: absorption column
density $N_{H}$, temperature $\kappa T_{bb}$ and peak bolometric flux  
 \be
  f_{0}=\frac{L_0}{4\pi d_L^2},
 \label{eq:norm}
 \ee
where $d_L$ is the luminosity distance to the source (see Section
\ref{s:pred}). We freeze the first parameter at $N_{H}=5\times
10^{20}$~cm$^{-2}$, which is close to the median value of Galactic
absorption for random positions in the extragalactic sky as calculated
by \cite{Esquej2008}. In addition to the Galactic absorption there might 
also be some intrinsic absorption due to edge-on orientation of the
accretion disk and/or screening by unbound stellar debris material
\citep{Strubbe2009}. However, in the former case the disk radiation
might be directed preferentially away from the observer, and so the
contribution of such events to the net detection rate is likely to be
small. In the latter case, the estimates of \cite{Strubbe2009}
indicate that unbound equatorial material builds up 
a 'vertical wall` of debris with huge column density in the radial
direction but subtending a small solid angle $\Delta\Omega\sim 0.1$~sr. 
Such screening will most likely result in complete
non-detection rather than detection of a TDE with a significant
excess of absorption relative to the Galactic value.
 
As follows from Eq. \ref{eq:temperature}, the second parameter of the
model obeys $\kappa T_{bb} \propto  M_{BH}^{-1/4}$. Unfortunately,
there are only crude $M_{BH}$ estimates for known TDE candidates. The
median observed temperature $\kappa T_{bb}\simeq 0.07$ keV
implies a SMBH with mass $M_{BH} \approx 10^6 M_{\odot}$. For
comparison, for $M_{BH}=5\times 10^6$ and $10^7 M_{\odot}$ one expects
$\kappa T_{bb}=0.047$ and $0.039$~keV, respectively. We shall use
these three fiducial values in our calculations.   
 
As the accretion rate of falling back material decreases, the
effective temperature of the radiation is also expected to decrease as 
$\kappa T_{bb}\propto\dot{M}^{1/4}$, as long as the thin disk
approximation remains valid \citep{Strubbe2009}. Nonetheless, late
observations of five TDE candidates from RASS \citep{Vaughan2004}
demonstrated no significant softening and in fact indicated a
marginal hardening for some of them, while the luminosities fell by
more than two orders of magnitude\footnote{It should be
    mentioned that the host galaxies of some of these candidates are
    classified as hosting an active nucleus.};
\cite{Komossa2004} found similar spectral hardening for
  another TDE candidate, RX J1242--1119. This behavior may have
resulted from transition to a different accretion mode accompanied by
Comptonization of disk radiation in a hot corona and/or significant
contribution from a relativistic jet. Taking into account these
observations and because we expect \e\ to detect TDEs mostly during an
early post-peak phase (see Section \ref{s:pred}), we disregard
possible spectral evolution of TDE-associated emission in our
estimates below.  
  
It is also worth mentioning that somewhat higher temperatures 
($\kappa T_{bb}\sim 0.12$~keV) have been reported for TDE candidates
detected in clusters of galaxies \citep{Cappelluti2009,Maksym2010}. It
is unclear whether this is owing to some selection bias, the specific
physical environment and/or details of the spectral analysis. In any
case, searches for TDE candidates with \e\ in galaxy clusters will be
challenging due to the strong diffuse X-ray emission from the hot
intracluster gas (which is much less of an issue for \textit{Chandra},
\citealt{Maksym2010}),  except for brightest flares in nearby
  clusters (like the one reported by \citealt{Cappelluti2009}), for
  which elaborate spectral analyses can be performed taking advantage of  
  \e's higher sensitivity below 0.3 keV compared to
  \textit{Chandra}. A further advantage in studying such events might
  be that galaxy clusters belong to the main
  scientific targets of eRASS \citep{Merloni2012} and will thus be
  extensively covered by follow-up observations.  
 
\section{Observational technique} 
\label{s:observ}

In formulating TDE identification criteria we should address the
\textit{completeness} of the resulting sample and \textit{reliability} 
of the selection procedure. As usual in such cases (see e.g. Chapter 7
of \citealt{Wall2003}), if we wish to maximize the number of
sources of the required type, we should be ready to deal with
significant contamination of the sample by misidentified sources. In
the case of eRASS, our primary goal is to obtain as large as possible
a sample of TDE candidates that will make it possible to accurately
measure the rate of such events in the local Universe. The associated 
systematic bias could then possibly be corrected by weighting the
candidates according to their reliability. Hence, we choose rather
loose identification criteria, based on key signatures of TDEs.  

\subsection{\e\ All-Sky Survey} 
\label{ss:eRASS}

The \emph{SRG} satellite will be rotating with a period of $T=4$~hours
around its axis pointed within a few degrees of the Sun, with the
telescopes observing the sky at right angles to the axis
\citep{Merloni2012}. Following the orbit of the L2 point around the
Sun the rotation axis will be moving at a speed of 1~deg per day. With
the 1~deg-diameter field of view (FoV), \e\ will complete an all-sky
survey every 6~months and a total of 8~scans will be performed over
the course of the mission. A typical position on the sky will receive
$\sim 240$~seconds of exposure during one scan. This exposure will be
achieved by 6 consecutive passages of a source through the \e\ FoV
separated by $\sim 4$-hour intervals (see \citealt{Khabibullin2012}
for further details). A typical visit of a point source will thus last
$\sim 1$~day, which is much shorter than the expected characteristic
timescale of the flux decay during a TDE flare (few months). Hence, we
shall regard such visits as single 240~s-long observations.

We note that in reality the eRASS exposure is not constant over the
sky and increases with ecliptic latitude. Most importantly, there are two 
areas of $\sim 1000$~deg$^2$ each, surrounding the ecliptic 
poles, that will be visited more than 20 times during each \e\ all-sky scan
\citep{Merloni2012}. Hence, the threshold for detecting TDE flares in
these regions will be $\sim 4$~times lower than the all-sky average
used in our analysis below. That will not only allow one to detect
weaker TDE flares (including events from more distant galaxies) 
but also to study in greater detail the X-ray light curves of bright TDE
flares. This can be particularly important for detecting TDEs during
their rise phase (See Section 4.2).

\subsection{Identification criteria} 
\label{ss:criter}

\begin{figure}
\centering
\includegraphics[width=\columnwidth]{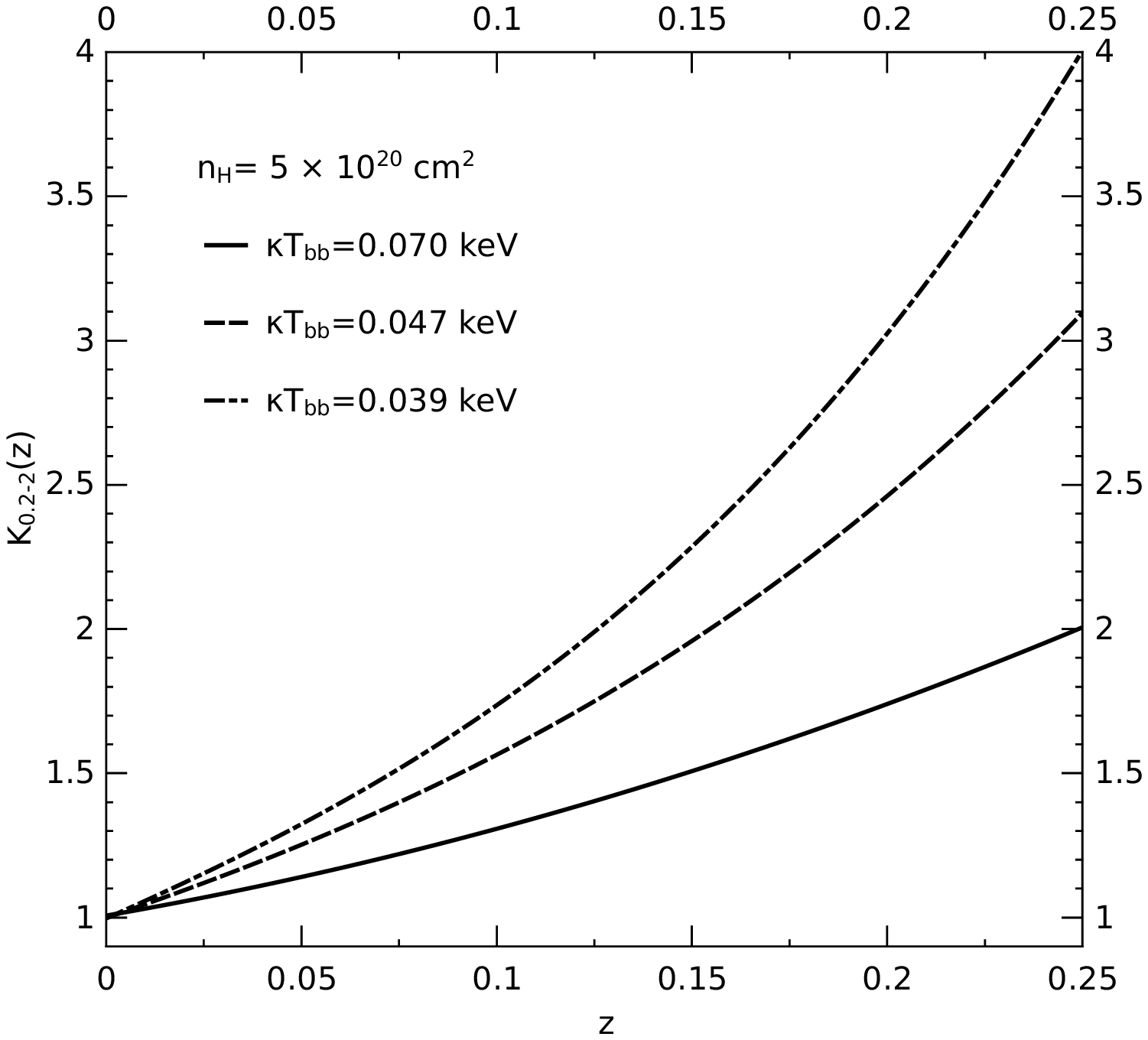}
\topcaption{Redshift spectral correction $K_{0.2-2}\doz$ for various
  spectral models representing spectra of TDE flares from
    SMBHs with $ M_{BH}=10^6M_{\sun} $  (black body with 
    $\kappa T_{bb}\simeq 0.070$ keV), $ M_{BH}=5\times 10^6 M_{\sun} $ 
  (black body with $\kappa T_{bb}\simeq 0.047$ keV), $ M_{BH}=10^7 M_{\sun} $
  (black body with $\kappa T_{bb}\simeq 0.039$ keV). All spectra were
    modified by Galactic absorption with $ N_H=5\times10^{20}$~cm$^2$.
} 
\label{fig:correction}
\end{figure}

For a TDE with a given spectral energy distribution, bolometric flux
$f$  (see Subsection \ref{ss:spectral}) and redshift $z$, the count
rate measured by \e\ between 0.2 and 2~keV can be written as  
\be
\label{eq:crate}
C_{0.2-2}(f,z)= \frac{A_{0.2-2}}{K_{0.2-2}\doz}f,
\ee
where $A_{0.2-2}$ (in units of counts~cm$^{2}$/erg) is the
flux-to-count conversion factor for a TDE at $z\to 0$ and the 
correcting factor $K_{0.2-2}\doz$ accounts for the cosmological
reddening of the spectrum. Both $A_{0.2-2}$ and $K_{0.2-2}$ depend on
the \e\ response matrix and TDE spectrum. We have computed the
redshift correction for different relevant spectral models using XSPEC
\citep{Dorman2001}, and it proves to be significant (see
Fig.~\ref{fig:correction}), especially for the model corresponding to
$M_{BH}=10^{7} M_{\odot}$. This has an obvious explanation: as
the intrinsic TDE spectrum becomes softer with increasing black hole mass,
a largely fraction of the photons red-shift out of the
\e\ energy band into softer bands (below 0.2~keV).  

Disregarding spectral evolution during the TDE, the count rate varies 
with time as $C_{0.2-2}(t_{observer})\varpropto
L\left(t_{source}(t_{observer})\right)$, where the relation 
$t_{source}(t_{observer})$ is governed by the source's redshift.
 
Additional information on the nature of TDE candidates
can be obtained by dividing the 0.2--2~keV energy band into several
sub-bands, e.g. 0.2--0.4, 0.4--1 and 1--2~keV. The first one is
sensitive to absorption along the line of sight, whereas the other two 
are well suited for estimating the effective spectral slope. We thus
introduce a softness ratio $SR=C_{0.4-1}/C_{1-2}$ to distinguish
'soft` and 'hard` sources (see Appendix for further details).  

Taking into account the anticipated TDE properties (see
Section~\ref{s:sign}) and eRASS characteristics (see
Section~\ref{ss:eRASS}), we propose to discriminate TDEs against
other types of X-ray transients to be detected during the survey (such
as AGN flares and GRB afterglows, see e.g. \citealt{Komossa2002})
based on the following two signatures: 
\begin{itemize}
\item large ($ \geq 10 $) variation between the count rates
  ($C_{0.2-2}$) measured in two consequent half-a-year scans (see
  Fig.~\ref{fig:obs}); 
\item soft spectrum (as indicated by the softness ratio $SR$),
  consistent with a weakly absorbed ($N_H <10^{21}$~cm$^{-2}$) power law
  with a photon index $\Gamma \geq 3$. 
\end{itemize}

In addition, since we are dealing with soft X-ray flares, it is
reasonable to limit TDE searches to Galactic latitudes $|b|>30^{\circ}$, 
where interstellar absorption $N_H<10^{21}$~cm$^{-2}$. That will also 
minimize contamination from cataclysmic variables and flaring dM stars, 
whose appearance can resemble TDEs (see \citealt{Donley2002}). 

\begin{figure}
\centering
\includegraphics[width=\columnwidth]{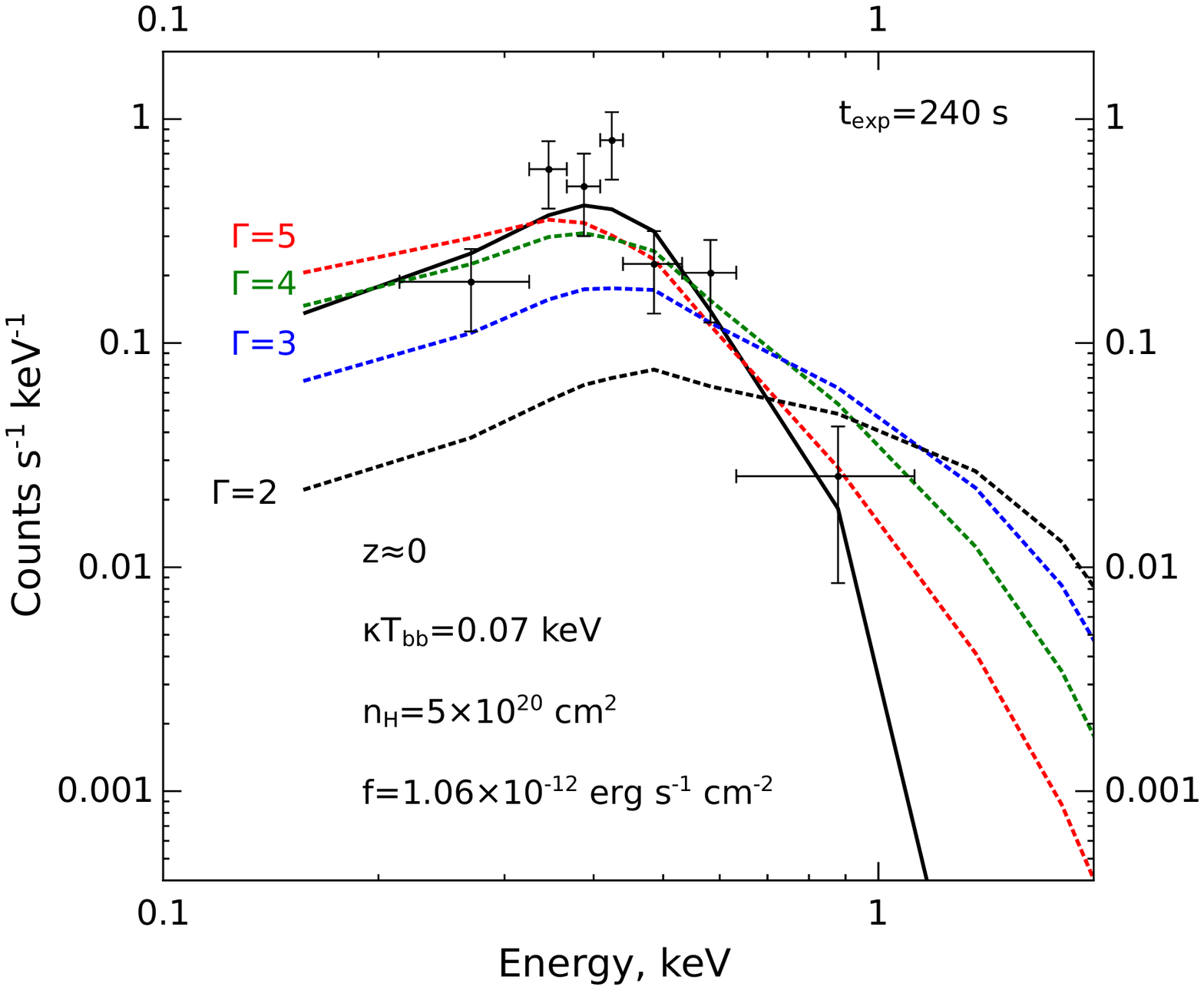}
\topcaption{Simulated \e\ spectrum of a TDE at $z\approx 0$ with flux
  near the eRASS detection threshold, i.e. providing 40~counts during
  240~s exposure time. The adopted TDE spectral model corresponds to
  $M_{BH}=10^6 M_{\odot}$ (see text) and the corresponding bolometric
  flux is $1.06\times10^{-12}$~erg~s$^{-1}$~cm$^{-2}$. The spectrum is
  binned in energy so that the significance of data points is
  $2\sigma$. Also shown are best fits of the data by different models:
  solid -- absorbed blackbody emission, dashed -- absorbed power laws
  ($N_H=5\times 10^{20}$~cm$^{-2}$) with the indicated slopes. 
  In this example, an  absorbed power law with $\Gamma\leq 3$ can be rejected with $ \approx $99\%  significance based on $\chi^{2}$ statistics with the weighting suggested by \citealt{Churazov1996} for spectral analysis in the case of low number of counts.}
\label{fig:spec}
\end{figure}

Although the peak X-ray flux during a TDE can greatly exceed the
corresponding quiescent flux, i.e. the flux produced by the TDE's
host galaxy, in reality the minimum flux will, in most cases, be
dominated by the \e\ background (see Appendix). Therefore, the first
of our requirements formulated above implies that a candidate TDE must 
be at least 10 times as bright as the background. Besides that, the
count rate must be high enough for the softness ratio $SR$ to indicate
that $\Gamma > 3$ rather than $\Gamma\sim 2$ as typical of AGN. As
demonstrated in Appendix, these conditions imply that at least
$N_{lim}=40$~counts (i.e. 20~counts inside the half power diameter
region) must be collected from a source between 0.2 and 2~keV during a
240~s exposure. The corresponding bolometric flux limits $f_{lim}$ for
various spectral models are given in Table \ref{t:sum}.  
Figure~\ref{fig:spec} shows a simulated \e\ spectrum (for
$M_{BH}=10^6M_{\odot}$) of a TDE with $f=f_{lim}$. Trial fits of this
spectrum by various characteristic models demonstrate that it is
feasible to reveal the spectral softness of a TDE using $\sim
40$~counts.  

We can finally formulate exact criteria that can be used for
identification of TDE candidates during eRASS:
\begin{enumerate} 
\item Extragalactic ($|b|>30^{\circ}$) location.
\item At least 40~counts (or 20~counts inside the half power diameter
  region) detected from the source during the $\sim 240$~s exposure in
  a given scan of the sky. 
\item At least a factor of 10 higher count rate in comparison with the
  preceding or succeeding scan. 
\item Softness ratio $SR\geq 2$ in the bright state. 
\end{enumerate} 

The last criterion should not be used in application to relativistic
TDEs, a subclass of TDEs discussed in Section~\ref{ss:jets} below.

\subsection{Detection in subsequent eRASS scans}
\label{ss:lcurve}

A TDE X-ray flare discovered in a given eRASS scan can remain
detectable by \e\ in its subsequent sky scans, i.e. 6 months, 12
months etc. after the TDE trigger. This provides an interesting
possibility of obtaining a crude (2 or more data points) long-term
light curve of TDE emission. Specifically, to obtain a significant
(see Appendix for details) X-ray detection from a continuing TDE
flare (whose position is known from its initial observation by \e),
just $\sim 4$~counts during the $\sim 240$~s exposure are
required. This implies, for instance, that a TDE flare will be
detectable by \e\ even if the X-ray flux has dropped by an order of
magnitude during 6 months after the trigger. Such late
detections, or even upper limits provided by non-detections, will be 
very useful for testing TDE models.    
  
\section{Predictions} 
\label{s:pred}

The intrinsic rate, $\mathcal{R}$, of tidal disruption events in the
local Universe, i.e. the number of TDEs per unit time and unit volume,
should depend on a number of factors, such as the mass distribution of 
SMBHs and the density profiles of nuclear stellar cusps,
which are fairly uncertain at present. Hence, it is difficult to
predict the TDE discovery rate for eRASS based on theoretical
grounds. However, we can try to estimate this rate using the existing
(poor) statistics of TDEs. Specifically, we can make use of an
estimate based on the detection of two candidates in the
\emph{XMM-Newton} Slew Survey and their non-detection in RASS:
$\mathcal{R}\simeq 5\times 10^{-6}$~yr$^{-1}$~Mpc$^{-3}$
\citep{Esquej2008}. Despite having a rather high uncertainty 
  and probably being affected by selection effects (see
  \citealt{Gezari2009}), this estimate is in line with other 
  observational results \citep{Maksym2010} and with some theoretical 
  expectations \citep{Wang2004}. 


\begin{table*}
\topcaption{Summary of the results. The adopted TDE intrinsic rate 
  $\mathcal{R}$ is a crude estimate based on the currently available 
  sample of X-ray selected TDEs (see text). It is assumed that all
  TDEs are associated with SMBHs of the same mass $M_{BH}$, for which
  three fiducial values are used. The detection limit is given in
  terms of the black body spectrum normalization $f=L_{b}/(4\pi
  d_{L}^2)$, where $L_{b}$ is the bolometric luminosity and $d_{L}$ is
  the luminosity distance to the source. The quantity $\mathcal{N}$
  gives the total number of TDEs with the 0.2--2~keV flux above the
  appropriate detection limit. It is not corrected for those events
  that were already detected in the preceding scans (see the next
  column for the relative fraction of such events). The next two
  columns provide the relative fractions of events that are triggered
  in the supercritical and rising phases of their light curves. The
  relative fraction of events for which \e\ will provide  at least
    2(or 3) significant data points for constructing a long-term ($\geq 6$
  months) X-ray light curve (see Section~\ref{ss:lcurve}) is given in the
  last column. 
  }
\centering
\begin{tabular}{ccccccccccc}\hline\hline
\multirow{2}*{Model} & {$\kappa T_{bb}$} & {$\mathcal{R}~/~10^{-5}$} &
         {$f_{lim}~/~10^{-12}$} &\multirow{2}*{$z_{lim}$}&{$\mathcal{N}$} &
\multicolumn{4}{c}{Fraction of events}\\ 
{}&{(keV)} &{(yr/~Mpc$^{3}$)} &{(erg/~s/~cm$^{2}$)}
&{}& {per scan} &in $\geq 2$ scans &  in Edd. phase &  in rising phase
& with  $\geq 2(3)$-point light curve \\
\hline
{$M_{BH}=10^6 M_{\odot}$} & {0.070} & {0.5} & {1.06} &{0.164}& {650}& {$\sim 1/12$}& {$\sim 1/2$}  & $\lesssim 4 \times 10^{-4}$ & {$\sim 4/5~(7/20)$} \\ 
{$M_{BH}=5 \times 10^6 M_{\odot}$} & {0.047} & {0.5} & {4.29} &{0.162}& {1240}& {$\sim 1/4$} &
{$\sim 1/2$}  &$\lesssim 1 \times 10^{-2}$ &  { $\sim 1~(16/20)$}\\ 
     {$M_{BH}=10^7 M_{\odot}$} & {0.039} & {0.5} & {10.81}
  &{0.142}& {1150}& {$\sim 1/3$}  & {$\sim 1/2$} & $\lesssim 5 \times 10^{-2}$ &{$\sim 1~(19/20)$} \\ 
\hline\\
\end{tabular}
\label{t:sum}
\end{table*}


\subsection{Detection rate}

We assume $\Lambda CDM$ cosmology with $\Omega_M=0.275,
\Omega_\Lambda=0.725$ and $h=0.7$ \citep{Komatsu2011}. The luminosity
distance as a function of redshift is then given by    
\be
\label{eq:dl}
d_L\doz=\oaz d_H ~\int_0^z\frac{dz'}{E\left( z'\right)},
\ee
where $d_{H}= \frac{c}{H_0}\approx 3.0 h^{-1}$~Gpc~$\approx 4.3$~Gpc is
the Hubble distance, and $E \doz
=\sqrt{\Omega_m\oaz^3+\Omega_\Lambda}$ \citep{Hogg1999}. Defining
$J\doz=\int_0^z\frac{dz'}{E\left( z'\right) }$, equation (\ref{eq:dl})
becomes $d_L\doz=d_H \oaz J\doz$.  

Given the flux detection limit $f_{lim}$ corresponding to a given
spectral model (Table~\ref{t:sum}), there exists the largest distance 
$d_{L,lim}$ (or the corresponding redshift $z_{lim}$) from which a TDE 
flare at its peak could be identified with \e:
\begin{equation} 
\label{eq:zlimimpl}
\frac{L_0}{4\pi d_L\dozsub{lim}^2 K\dozsub{lim}}=f_{lim}  
\end{equation}
or
\begin{equation}
\label{eq:zlim}
\oazsub{lim}^2J\dozsub{lim}^2 K\dozsub{lim}= \frac{L_0}{4\pi
  d_{H}^2 f_{lim}}.  
\end{equation}

Solving this equation numerically with $L_0=1.3 \times
10^{44}$~erg~s$^{-1}$ (the Eddington luminosity for a black hole with
mass $M_{BH}=10^6 M_{\odot}$ and geometrical dilution factor
$\eta=1$), we obtain $z_{lim}=0.164$. This is approximately two times 
further than the distance within which TDEs could be found with RASS
($z\approx 0.09$, \citealt{Donley2002,Esquej2008}).  
Similar calculations for $M_{BH}=5\times 10^6$ and $10^7M_{\odot}$,
again assuming the Eddington luminosity at the peak phase, 
result in $z_{lim}=0.162$ and 0.142, respectively. Remarkably, the
limiting redshifts for the different SMBH masses prove to be 
similar. This happens because although the peak TDE bolometric
luminosity increases with $M_{BH}$, the spectrum becomes softer and
most of the emission appears at frequencies below the \e\ energy 
range.  

Consider now a small volume of the Universe at redshift $z$ with
radial width $dz$, covering an opening angle $\omega$ on the
observer's sky. The comoving volume of the slice is
\begin{equation}
\label{eq:dV}
	dV_{c}\doz= \omega d_{H}^{3} \frac{J\doz^{2}}{E \doz} dz	
\end{equation}
(\citealt{Hogg1999}; see also Chapter~13 of
\citealt{Peebles1993}). Neglecting any evolutionary changes in the
galaxy population between $z_{lim}$ and $z=0$, we can calculate the
rate of TDEs in this volume as $d\mathcal{N}\doz=\mathcal{R}dV_{c}\doz dt$.

Taking into account both the maximum and decay phases of the light
curve (eq.~(\ref{eq:decay})), a TDE flare remains detectable
during a comoving time
\be 
\label{eq:utime}
\tau_{lim}\doz= \tau\edd \left(\frac{L_0}{ 4\pi d_H^2 f_{lim}}
\frac{1}{\oaz ^2 J\doz ^2 K\doz } \right)^{3/5}.
\ee
Hence, at a given moment of observation there are 
\begin{equation}
\label{eq:dN}
	d\mathcal{N}\doz= \mathcal{R}\times\tau_{lim} \doz\times dV_{c}\doz 
\end{equation}
detectable TDEs. Substituting the corresponding expressions for
$dV_{c}\doz$ and $\tau_{lim} \doz$ and integrating from $z=0$ to
$z_{lim}$, we obtain the total number of detectable TDEs:
\begin{equation}
\label{eq:N}
	\mathcal{N}=\alpha \times I(z_{lim}),
\end{equation} 
where 
\be
\label{eq:alpha}
\alpha=\omega \mathcal{R} d_{H}^{3} \tau\edd \left( \frac{L_0 }{4\pi
  d_H^2 f_{lim}}\right)^{3/5} 
\ee
and
\be
\label{eq:integ}
I(z_{lim})= \int_0^{z_{lim}} \frac{K\doz^{-3/5} J\doz^{4/5}}{\oaz^{6/5}~ E\doz} dz.
\ee 
Finally, expressed in convenient units, the total number of detectable
TDEs at a given moment is 
\be
\label{eq:rate}
\mathcal{N} \simeq  9.51 \times 10^4 \left( \frac{\omega}{2\pi}\right)
\left( \frac{\mathcal{R}}{10^{-5}}\right)
\left ( \frac{\tau\edd}{0.1} \right)
\left ( \frac{L_0}{10^{44}}\right)^{3/5} 
\left ( \frac{f_{lim}}{10^{-12}}\right)^{-3/5}I\dozsub{lim}.
\ee

\subsection{Dependence on M$_{BH}$ and the phase of the TDE light curve}
\label{ss:depend}

Estimates of the SMBH mass function \citep{Hopkins2007,Greene2007} 
indicate that the spatial density of SMBHs is roughly constant in the 
mass range from $\sim 10^6$ to $10^7 M_{\odot}$ but 
depends on redshift. The TDE rate may demonstrate some correlation 
\citep{Brockamp2011} or anti-correlation \citep{Wang2004} with the
central SMBH mass. Therefore, since the maximum TDE distance for eRASS 
is almost independent of SMBH mass, we may expect the TDEs found
during eRASS to be distributed over the $M_{BH}$ range from $\sim
10^6$ to $\sim 10^7 M_{\odot}$. Since this distribution is currently
unknown, we consider three different scenarios in which all TDEs
detected during eRASS are characterized by a fixed $M_{BH}$ value:
$10^6$, $5\times 10^6$ or $10^7 M_\odot$. The resulting estimates are
summarized in Table~\ref{t:sum}.   

\begin{figure}
\centering
\includegraphics[width=\columnwidth]{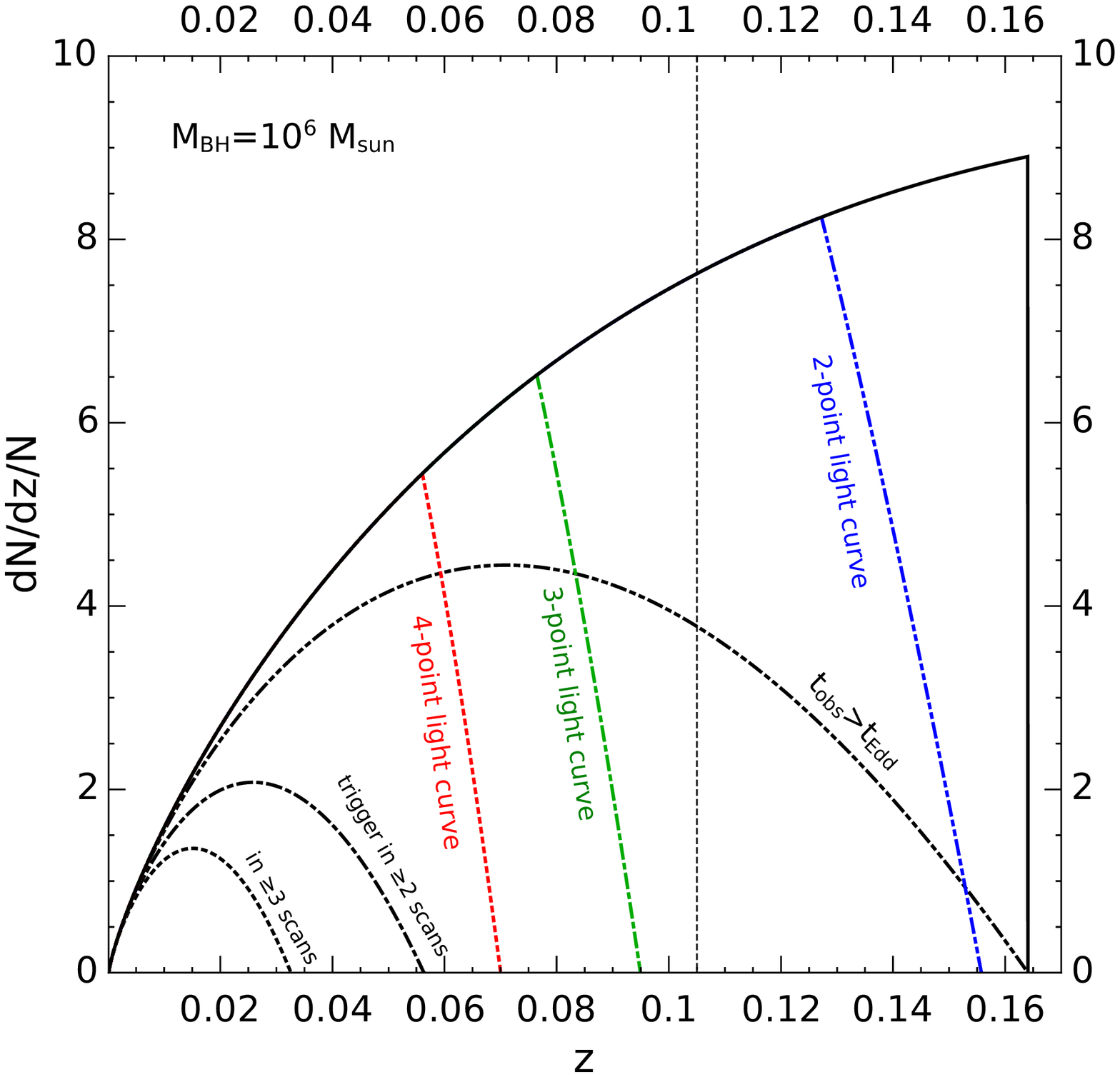}
\topcaption{Distribution of TDE candidates over redshift for
  $M_{BH}=10^6 M_{\odot}$. The total sample is shown by the solid black
  line. The events triggered as TDE candidates during the declining
  (post-supercritical) phase of the X-ray flare are shown by the
  two-dot-dashed black line. The events independently triggered as TDE
  candidates in more than one (two) eRASS scans are shown by the
  dot-dashed (dotted) black line. The blue, green and red lines
  denote events for which at least 2-,3- and 4-point X-ray light curve can
  be obtained using data from consecutive eRASS 
  scans. The thin vertical dashed line marks the redshift at which
  the numbers of events triggered during the supercritical and
  declining phases are equal.}
\label{fig:times}
\end{figure}

\begin{figure}
\centering
\includegraphics[width=\columnwidth]{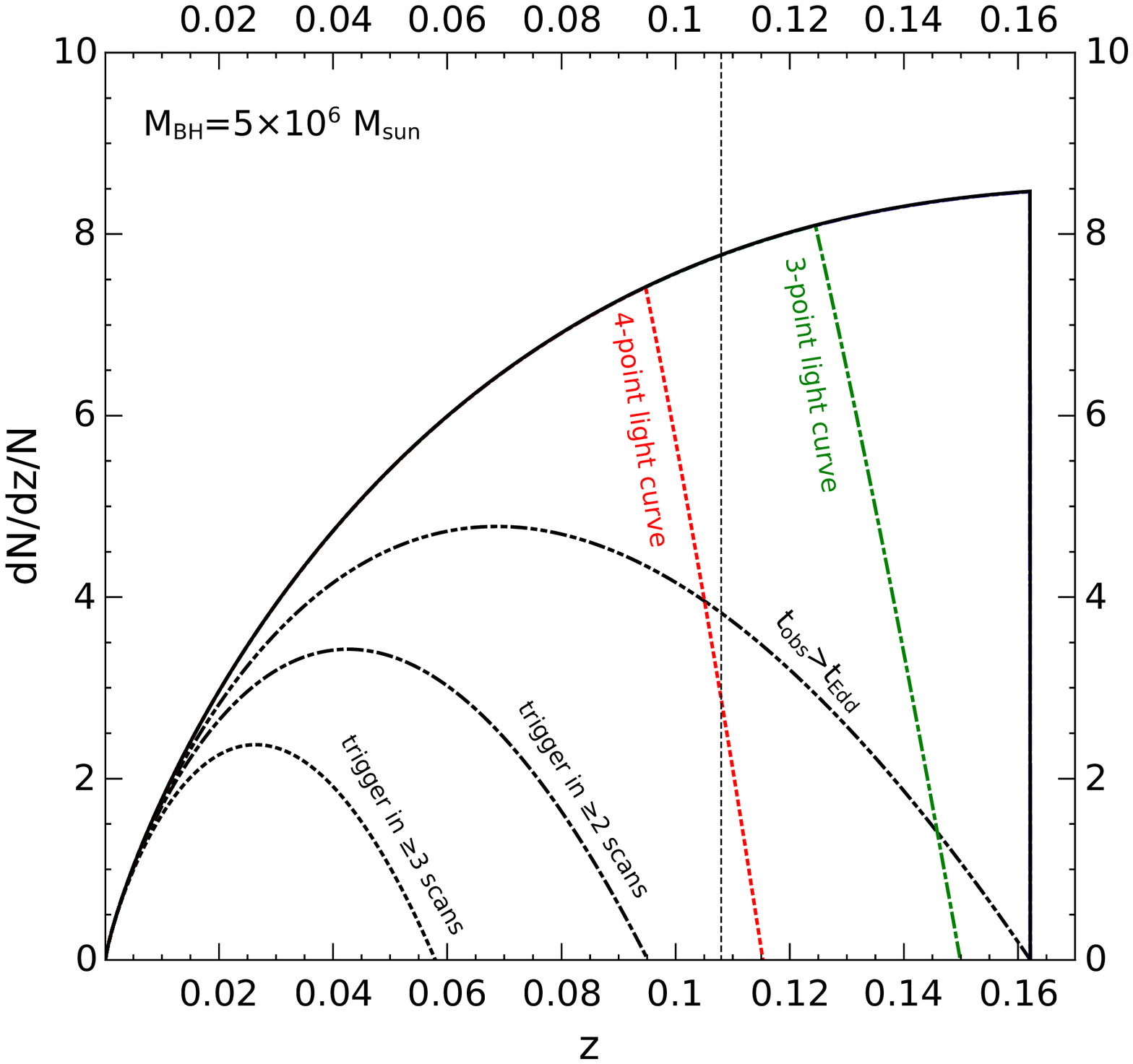}
\topcaption{Same as Fig.~\ref{fig:times} but for $M_{BH}=5\times
  10^6 M_{\odot}$.}  
\label{fig:times5}
\end{figure}

\begin{figure}
\centering
\includegraphics[width=\columnwidth]{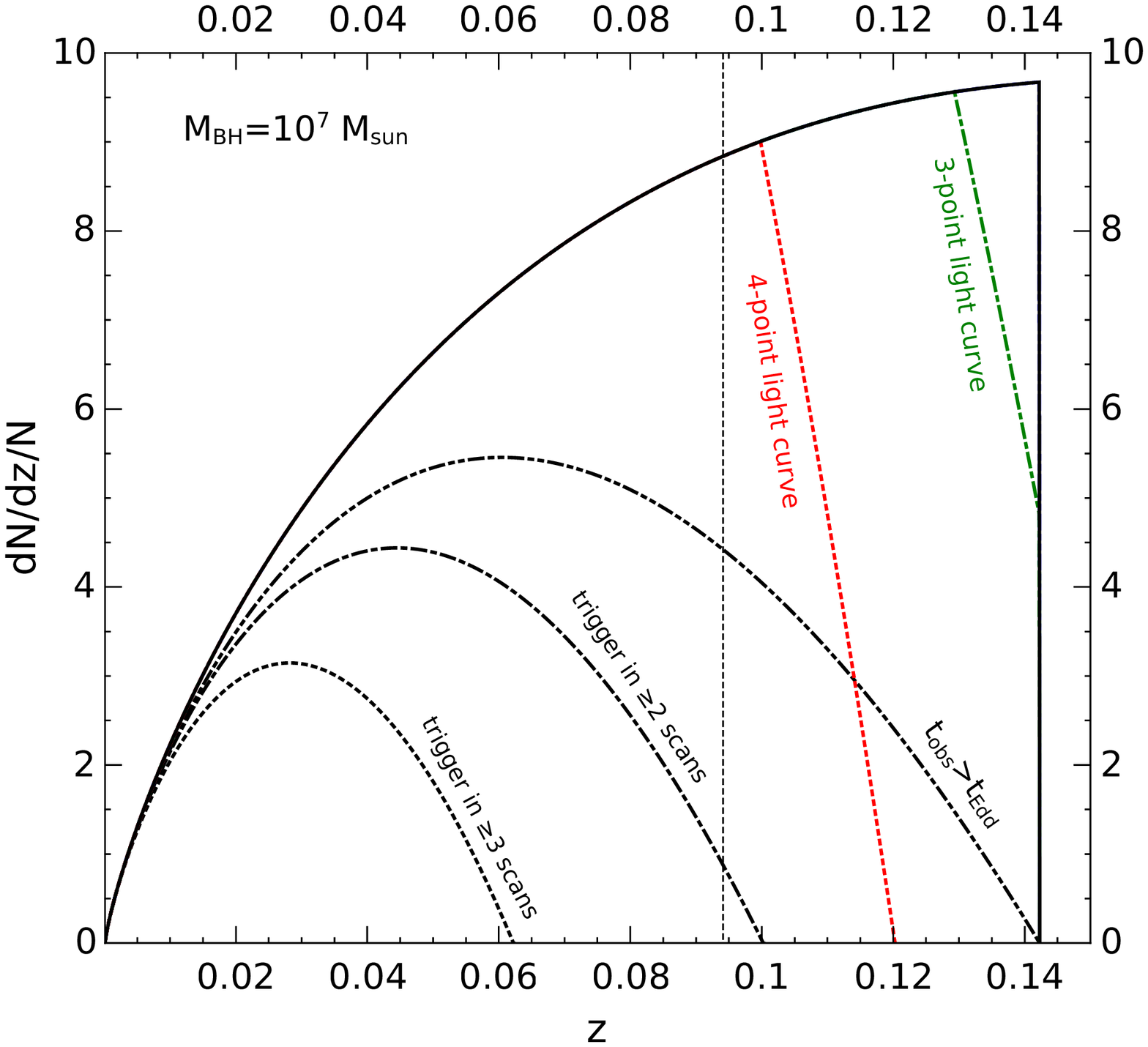}
\topcaption{Same as Fig.~\ref{fig:times} but for $M_{BH}=10^7 M_{\odot}$.} 
\label{fig:times10}
\end{figure}

For $M_{BH}=10^6 M_\odot$ and adopting $\mathcal{R}=5\times
10^{-6}$~yr$^{-1}$~Mpc$^{-3}$, $\omega=2\pi$ (which corresponds to
the $|b|>30^\circ$~sky) and $R_{p}=3R_{S}$, we find from
equations~(\ref{eq:tedd}) and (\ref{eq:integ}) that
$\tau_{Edd}=0.1$~yr and
$I\dozsub{lim}\approx\frac{1}{4.5}z_{lim}^{1.5}=0.0153$, and finally 
from equation~(\ref{eq:rate}) that $\mathcal{N}\approx 
650$ events can be discovered in the second eRASS scan through
comparison with the first one. Figure~\ref{fig:times} shows  
the distribution of the  TDE flares over redshift and as a
function of time passed after stellar destruction. Approximately half
of the  TDE triggers are expected to take place during the supercritical
accretion phase. For $\sim 1/12$ of the events, a second  trigger
(during the subsequent eRASS scan) will be possible, and in  
rare cases, three or more independent triggers will be
possible. Thus, the majority of TDEs will be
`new' for each scan, i.e. not seen during the preceding one. 
For $M_{BH}= 5\times 10^6$ and $10^7 M_{\odot}$, the fraction of `old'
 TDEs, i.e. events observed in more than one eRASS scan is
higher (Fig.~\ref{fig:times5} and Fig.~\ref{fig:times10}) since the
characteristic decay time positively correlates with the black hole
mass (see Section~\ref{s:sign}).  

Besides that, it turns out that the majority of TDEs discovered in a
given eRASS scan will still be detectable  (i.e. significantly
  exceeding the background flux level) by \e\ (see
Section~\ref{ss:lcurve}) in the subsequent sky scan(s), i.e. 6 months,
12 months etc. after the initial trigger. Hence, \e\ will typically
provide 2--4 significant data points for constructing a long-term
X-ray light curve of a TDE flare (see Table \ref{t:sum} and
Figs.~\ref{fig:times}--\ref{fig:times10}). As a result, eRASS will
provide a unique sample of TDE light curves, which can be used to test
TDE models. 

The above estimates were based on the assumption that the peak
accretion rate exceeds the Eddington limit. In reality, only TDEs
caused by sufficiently close stellar passages can provide a
supercritical accretion rate, whereas more distant disruptions may be
characterized by subcritical maximum accretion rates. Since the
Eddington limit increases proportionally to the SMBH mass, the
fraction of sub-Eddington TDEs may become significant for high
$M_{BH}$; e.g. \cite{Ulmer1999} estimate that the peak accretion rate  
exceeds the Eddington limit in approximately 100\% of events 
for $M_{BH}\sim 10^6 M_{\odot}$ and in $\sim 50$\% for 
$M_{BH}=10^7 M_{\odot}$. In such a case, our estimates for the number
of TDEs detected during eRASS should be reduced accordingly.

As was mentioned in Section \ref{ss:temporal}, there is a
non-negligible chance of detecting some TDEs during the rising
phase of the X-ray flare, especially for large $M_{BH}$ and/or
$R_p$. We may place a rough upper limit on the number of such 
  triggers: $\mathcal{N}_{rise}/\mathcal{N}_{Edd} \lesssim
\tau_{i}/\tau_{Edd}$, where $\mathcal{N}_{Edd}$ is the number of TDEs
detected during the supercritical phase. For $M_{BH}=10^7 M_{\odot}$,
up to $\sim 20$ rising TDE flares can appear in each eRASS scan (see
Table~\ref{t:sum}). Since the rising phase is expected to last between
a fraction of an hour and a few days, it might in fact be possible to
see how the X-ray transient brightens up between its $\gtrsim 6$
consecutive passages through the \e\ FoV every $\sim 4$ hours. 

As we noted before, there are two regions in the sky that will
  receive significantly increased exposure during eRASS, with more
  than 20 consecutive passages of a given source 
  through the \e's FOV. These regions are centered around the Northern
  and Southern ecliptic poles (NEP and SEP), with $b\simeq 30^\circ$
  and $b\simeq -30^\circ$, respectively, and so, unfortunately, about
  half of the area of these 'deep fields' proves to be outside our
  proposed   $|b|>30^{\circ}$ TDE search region. 
Furthemore, the 'extragalactic', i.e. $|b|>30^{\circ}$, part of the SEP
region contains the Large Magellanic Cloud, which will probably also
complicate detection of TDE   flares there. As regards the NEP region,
TDE flares of given luminosity could be detected from $\sim 1.5$ times
larger distances ($ z_{lim}$) than elsewhere, as a result of the
4-fold increase in sensitivity (see Eq. \ref{eq:zlim}). Therefore,
according to Equations (\ref{eq:alpha}, \ref{eq:rate}), one might
expect a $ \sim 4 $ times higher TDE rate (per unit solid angle) in
the NEP region compared to the average over the sky.

\subsection{Revealing TDEs through eRASS--RASS comparison}
\label{ss:comparison}

Apart from comparing the results of successive eRASS sky scans, TDEs
can also be discovered by cross-correlating the data of the first
eRASS scan with RASS data\footnote{The sky coverage of
    \textit{Chandra} and \textit{XMM-Newton} serendipitous surveys are 
    too small for this purpose (see \citealt{Brandt2005}), while
    the \textit{XMM-Newton} Slew Survey is similar in sensitivity to
    RASS (below 2~keV) but effectively covers only about a third of the sky
    \citep{Esquej2008}(see also \cite{Warwick2012} for a more recent reference).}. The characteristic unabsorbed 0.5--2 
keV flux limit for RASS point sources is $3\times 
10^{-13}$~erg~s$^{-1}$~cm$^{-2}$ \citep{Brandt2005}. Hence, to detect
a factor of 10 increase in flux, a candidate TDE must reach a
0.5--2~keV flux $>3\times 10^{-12}$~erg~s$^{-1}$~cm$^{-2}$  during
the first eRASS scan. This unabsorbed flux corresponds to a bolometric
flux $f_{lim}\simeq 4.6\times 10 ^{-12}$~erg~s$^{-1}$~cm$^{-2}$ for
the spectral model corresponding to $M_{BH}= 10^6 M_{\odot}$). This
limit is approximately 4 times higher than for TDE detection
based on eRASS scan-to-scan comparison (see Table~\ref{t:sum}). Solving
equation (\ref{eq:zlim}) for $f_{lim}$ results in $z_{lim}\simeq
0.095$. Using equation (\ref{eq:rate}), we find that cross-correlation
of the first eRASS scan with RASS will provide an additional 
$\sim 125$ TDE candidates (assuming $\mathcal{R} =5\times
10^{-6}$~yr$^{-1}$~Mpc$^{-3}$, $\omega =2\pi$ and $M_{BH}= 10^6
M_{\odot}$.

In addition, $\sim 15$ TDE candidates could be found using the method of
\cite{Donley2002}, i.e. by looking for \emph{ROSAT} sources with soft
X-ray spectra that were at least $\gtrsim 10$ times brighter during RASS
than in subsequent observations. In their study \citep{Donley2002}
used \textit{ROSAT} PSPC data, which cover only $\sim 9$\% of the sky,
for the follow-up, whereas with eRASS the analysis can be done for
50\% of the sky ($\vert b \vert \geq 30^{\circ}$). Follow-up 
radio observations of such candidates could be useful for constraining 
models of TDE-associated jets (\cite{Bower2012}; see Subsection
\ref{ss:followup}). 

\subsection{Tidal disruption events with jets}
\label{ss:jets}

Only two TDE flares with signatures of a relativistic jet
have been discovered so far \citep{Cenko2012}. Moreover, these
detections were triggered in the hard X-ray (15-50 keV) band 
by the Burst Alert Telescope (BAT) on \textit{Swift}. Therefore, it is 
difficult to predict how many such transients can be found during
eRASS.  

A very rough estimate can be obtained by assuming that all TDEs with
jets are similar to Sw~J2058+05, the brightest \textit{Swift} event of
this type. Its estimated isotropic X-ray (0.3--10~keV) luminosity is 
$L_{X,iso}=3\times 10^{47} $ erg s$^{-1}$ and the spectrum is  
consistent with a power law with $\Gamma\simeq1.6$ absorbed by
a column density $N_{H}\simeq 2.6 \times 10^{21}$~cm$^{-2}$, intrinsic
to the source located at $z=1.185$ \citep{Cenko2012}. We simulated an
\e\ observation of a source with such a spectrum at $z=1$  and
found that our count rate detection limit corresponds to an unabsorbed
flux $F_{0.3-10}=2.3\times 10^{-12}$~erg~s$^{-1}$~cm$^{-2}$ (we 
disregarded the K-correction since the spectral shape above 10~keV is  
unclear, whereas the spectrum below 10~keV is variable and corresponds
to $K\simeq const$ for $z\gtrsim 1$). Assuming a peak isotropic
X-ray (0.3--10~keV) luminosity\footnote{This adopted value is somewhat
  higher than the luminosity measured for Sw~J2058+05, but the
  corresponding \textit{Swift}/XRT observations started some 10 days
  after the beginning of the hard X-ray flare and so may have missed
  the brightest phase of the event.} $L_{X,iso}\sim 5\times
10^{47}$~erg~s$^{-1}$, the limiting 
redshift is $z_{lim}\simeq 4.5$. Therefore, TDEs with jets can
potentially be found with eRASS even in the distant Universe. Clearly,
the actual number of such events will strongly depend on the cosmic
evolution of galaxies and their nuclei.   
Ignoring the redshift dependence, the existing \textit{Swift}
statistics implies that roughly $\sim 1$ TDE with jets may be detected
during each eRASS scan. 

An independent upper limit on the rate of TDEs with jets is provided
by \textit{ROSAT} data. The analysis of \cite{Donley2002} revealed
only one high-amplitude ($\geqslant 20$) flare with a hard spectrum
($\Gamma < 2$). This event is associated with the Seyfert~2 galaxy
SBS~1620+545 \citep{Carrasco1998} at $z=0.0516$. On the \textit{WISE}
IR color--color diagram \citep{Wright2010}, the object falls on the 
boundary between Seyferts and spiral galaxies ([W1-W2]=0.6,
[W3-W2]=2.9, Data Tag: ADS/IRSA.Gator$ \# $2012/1017/032431$ \_
$5650). The observed-frame absorption-corrected
0.2--2.4~keV luminosity is $\sim 10^{43}$~erg~s$^{-1}$ in the bright
state \citep{Donley2002}, which is much less than the observed
luminosity of relativistic TDEs. Therefore, it seems more likely that
this hard flare is an example of extreme variability in Seyfert~2
galaxies. Observation of SBS~1620+545 during eRASS will probably
clarify the situation, since there have been no post-\textit{ROSAT} X-ray
observations of the source. Given that the same analysis of RASS data
\citep{Donley2002} revealed 5 soft flares -- TDE
candidates, we can tentatively conclude that the rate of relativistic
TDEs is at least 5~times less than for `ordinary' TDEs. Hence, perhaps
not more than $\sim 150$ relativistic TDEs will be detected during
each eRASS scan.

In order to search for relativistic TDEs in eRASS data, one could use
the same count rate criteria as for 'normal' TDEs
(see Section~\ref{ss:criter}) but the spectral softness criterion
should obviously be omitted or modified so that it selects hard
rather than soft X-ray flares.

\section{Discussion}
\label{s:discussion}

In order to reveal the true nature of the numerous TDE candidates 
to be found during eRASS, some additional identification/follow-up work 
should be done, as we discuss below.

\subsection{Cross-correlation with other surveys}
\label{ss:correlation}

The localization accuracy for point sources detected during eRASS is
determined by the \e\ point spread function (PSF) averaged over its 
FoV ($\simeq \varnothing 1^{\circ} $). For the faintest sources, it is
expected to be $\sigma \simeq 12''$ (corresponding to a half-power
diameter, HPD, of $\simeq 29 ''$,
\citealt{Merloni2012}).  Localization of bright sources,
  such as TDE candidates, can be significantly better since the
  centroid determination error is inversely proportional to the square
  root of the number of counts detected. Further improvement for TDE
  flares could be achieved using the fact that 
  several of the total $\gtrsim 40$ counts will be detected in the
  on-axis region of the telescope's FoV, where the HPD is $\sim 15'' $
  \citep{Merloni2012}. We may thus conservatively estimate that
  \e\ should be able to localize TDE flares to better than 10''. At
  redshifts typical of the eRASS TDE sample, $z\simeq 0.1$, the
  corresponding linear size is thus $<20$~kpc. Such accuracy
should be sufficient for searching for TDE host galaxies in optical
(e.g. SDSS\footnote{http://www.sdss3.org/}, \citealt{SDSS2012}, and
Pan-STARRS\footnote{http://pan-starrs.ifa.hawaii.edu/public/home.html},
\citealt{Kaiser2010}) and infrared
(e.g. \textit{WISE}\footnote{Wide-field Infrared Survey Explorer,
  http://irsa.ipac.caltech.edu/Missions/wise.html},
\citealt{Wright2010}) catalogs. However, association of a TDE
candidate with the nuclear region of a galaxy will be possible only
for bright flares and/or nearby galaxies.

Should the redshift of the likely host galaxy of a TDE candidate be
known, it may immediately indicate a very high luminosity of the
transient and hence exclude its extranuclear origin (i.e. a
highly variable ultraluminous X-ray source). To additionally exclude
soft X-ray flares from active galactic nuclei, an optical spectrum
\citep{Veilleux1987} and IR colors \citep{Wright2010} of the
counterpart galaxy could be used, if 
available\footnote{Such flares, if found by eRASS, may be interesting
  themselves for studying AGN variability mechanisms.}.
Therefore, cross-correlation efforts can greatly help in
confirming/rejecting the TDE nature of candidates found
during eRASS. 

\subsection{Follow-up observations}
\label{ss:followup}
The expected rate of 'TDE triggers' (see Section~\ref{s:pred}) during
eRASS implies that there will be roughly two TDE alarms every day
(data transfer from the spacecraft is planned to occur once a day). For 
some (perhaps a minority) of them, host galaxy identification will not
be immediately clear. Optical follow-up observations for such events
could clarify the situation and possibly also detect an optical signal
associated with the TDE itself if carried out early enough after the
TDE peak. Given the spectral energy distribution of TDE flares, it is
more likely, however, to detect fading emission in X-ray
(e.g. with \textit{Chandra} or \textit{XMM-Newton}) and/or UV
(\textit{GALEX}\footnote{Galaxy Evolution Explorer,
  http://www.galex.caltech.edu/ }) follow-up observations. Moreover,
\textit{Chandra} can provide excellent localization ($\lesssim 1''$)
and a high quality spectrum in the 0.3--7.0~keV energy band after
a few kilosecond observation (see, e.g., \citealt{Vaughan2004}). An
even higher signal-to-noise spectrum can be provided by
\textit{XMM-Newton}. A complementary UV detection by \emph{GALEX}
would help constrain TDE emission models even better than with X-ray
data alone \citep{Gezari2009}.  

In addition, radio follow-up observations may be crucial for
identification of relativistic TDEs, as has been the case for both
events discovered by \emph{Swift}
\citep{Metzger2012,Giannios2011,vanVelzen2011B}. On the contrary, 
searches for radio emission accompanying `ordinary' TDEs have not yet
given conclusive results \citep{Bower2012,vanVelzen2012}. 

We further note in this connection that blazar flares could in
principle resemble relativistic TDEs. Indeed, blazar emission is also
believed to originate in the vicinity of a SMBH from a relativistic
outflow directed at the observer. However, the two relativistic TDEs
discovered by \textit{Swift} were clearly different from known 
blazar flares and from any other known types of transients in that i)
the X-ray luminosity was very high and ii) the 
broad-band spectral energy distributions (from radio to hard X-ray)
were inconsistent with the so-called blazar sequence, in particular
the optical to X-ray luminosity ratio was very low (see Figs.~4--6 in 
\citealt{Cenko2012}). Therefore, follow-up observations will likely be
able to discriminate relativistic TDEs from blazar flares among hard
X-ray flares detected by eROSITA. We also note that identification of
new blazars in eRASS data is an important scientific task by itself,
which can be addressed by different known methods, e.g. using infrared
colors provided by the \textit{WISE} all-sky survey
\citep{Massaro2011}.

\section{Conclusions}
\label{s:conc}

We have demonstrated that eRASS will be very effective at finding
stellar tidal disruption events. A uniquely large sample of several
thousand TDE candidates can be obtained, making it possible to
accurately measure the rate of stellar disruptions by SMBHs with
masses from $\sim 10^6$ to $\sim 10^7M_{\odot}$ and thus to obtain 
a census of dormant SMBHs and their associated nuclear stellar cusps
in the local Universe (within $z\sim 0.15$). Roughly half of the TDEs
will be observed by \e\ during the supercritical (Eddington) phase of
accretion and the rest during the subsequent decay. In addition, there
might be up to $\sim 10^2$ TDE flares caught during the rising phase,
i.e. when the debris of the disrupted star have just reached the black
hole. Furthermore, for the majority of the TDE candidates, \e\ will
still be detecting their fading X-ray emission in 1--3 subsequent sky
scans, i.e. 6--18~months after 
the initial trigger. Hence, a unique sample of TDE light curves will
be obtained, diminishing the need for dedicated X-ray follow-up
efforts. There will thus be plenty 
of information on various stages of accretion of stellar material by
SMBHs, setting stringent constraints on TDE theory.

Importantly, TDE candidates can be recognized as such almost
immediately (within a day) after their detection by \e, so that prompt 
multiwavelength follow-up observations can be organized. Such
dedicated efforts may lead to the discovery of peculiar events among
the many TDEs discovered by eRASS, e.g. associated with a rapidly
rotating SMBH or with the disruption of a rare type of star or a
planet (such as the recently discovered hard X-ray transient 
IGR~J12580+0134, \citealt{Nikolajuk2013}).

Of particular interest is the possibility of obtaining a substantial
sample of relativistic TDEs. Due to their extreme power, such
transients can be detected by eRASS up to great distances ($z\sim
4$). The increased sample can shed light on the relation of
jet-dominated TDE flares to 'normal' ones. Detection of
relativistic TDEs at large redshifts will also provide a unique
opportunity to study quiescent SMBHs (i.e. not in quasars) in the
distant Universe.      

\section*{Appendix}
\label{s:app}

Since we consider only inactive galaxies as possible hosts of TDEs,
the \e\ count rate associated with their quiescent (non-flaring) state
will likely be below the background level, which is estimated at
$3.74\times 10^{-3}$~cts~s$^{-1}$~arcmin$^{-2}$ in the 0.2--2~keV
energy range \citep{Merloni2012}\footnote{This estimate includes the
  contributions of the cosmic X-ray background (CXB) and particle 
  background, with the former dominating below
  2~keV. Below 0.5~keV, a major contributor to the CXB is diffuse
  background emission from the Milky Way, whose intensity
  significantly ($\gtrsim 35 $\%) varies over the sky
  \citep{Lumb2002}. This effect could be included in the real data
  analysis using the \emph{ROSAT} map of the Local Hot Bubble emission
  \citep{Snowden1997}.}. This corresponds to 0.165~counts during a
240~s exposure inside the region of half-power diameter (HPD, 29'')
averaged over the FoV. Assuming Poisson distribution of the background
counts, the probability for $>2$ photons to be detected during this time
inside HPD is $ P_{bg}=0.00066$, i.e. $<10^{-3}$. Since we are looking
for $\gtrsim 10$~times amplitude flares above the background level,
the detection limit for the source in the bright phase corresponds to
$2\times 10=20$~counts inside the HPD region during 240~s. Therefore,
the source count rate must exceed $C_{lim}= 0.167$~cts~s$^{-1}$ in the
0.2--2~keV energy band. The false rejection probability of $\sim
10^{-3}$ is sufficient for our purposes, since the resulting sample of
TDEs is not expected to exceed a few thousand events. 
  
Having 20 photons with energies from 0.2 to 2~keV inside the HPD
region, it should also be possible to distinguish a soft X-ray
spectrum as expected for a TDE flare from a harder spectrum typical of
AGN. This distinction can be done in terms of a softness ratio,
$SR=C_{0.4-1}/C_{1-2}$, the ratio of the count rates in the 0.4--1~keV
and 1--2~keV energy bands. This ratio depends on the absorption column
density $N_H$ only weakly: for a power law spectrum with $\Gamma=2$,
$SR(N_H=10^{20}~{\rm cm}^{-2})\simeq 1.9$ and $SR(N_H=5\times
10^{20}~{\rm cm}^{-2})\simeq 1.6$; for $\Gamma=3$,
$SR(N_H=10^{20}~{\rm cm}^{-2})\simeq 4.0$ and $SR(N_H=5\times
10^{20}~{\rm cm}^{-2})\simeq 3.3$. Hence, having 20 counts within the
HPD region and imposing the condition $SR>2$, one can exclude at least
$\sim 50$\% of 'hard` ($\Gamma=2$) sources but retain almost all 
`soft' sources: for $\Gamma=3$ and $N_H=5\times 10^{20}~{\rm
  cm}^{-2}$, $SR>2$ with $\simeq 80\%$ probability. Since a typical 
TDE signal is believed to be even softer ($\Gamma\simeq 5$ in the
0.4--2~keV energy band), only a tiny fraction of true TDEs will
be missed if the criterion $SR>2$ is used. In fact, one could
use an extraction region several times larger than the HPD region if
the accurate shape of the PSF is known. Combining counts with
appropriate weights could increase the signal-to-noise ratio by a
factor of $\sim \sqrt{2}$ and also the efficiency
of the softness ratio criterion ($\sim 90\%$ of sources with
$\Gamma=3$ and $N_H=5\times 10^{20}~{\rm   cm}^{-2}$ will be
retained). 

We also point out the possibility of exploiting the \e\ sensitivity in
the 2--10~keV energy range. For a source near the detection
threshold,  a `soft' ($\Gamma=3$ and $N_H=5\times
10^{20}$~cm$^{-2}$) spectrum will provide 0.30~cts in the HPD
region in the 2--10~keV energy band for a 240~s exposure, whereas a
`hard' ($\Gamma=2$ and $N_H=5\times 10^{20}$~cm$^{-2}$) spectrum will
provide 1.12~cts. The hard (2--10~keV) \e\ background is
estimated at 0.04~cts in 240~s inside the HPD region
\citep{Merloni2012}, although this number is rather uncertain due to
the lack of data about the high energy particle environment near the
L2 point, which is believed to determine the \e\ background in this
energy range. Thus, a non-zero flux in the 2--10~keV range is expected
for $\gtrsim 1/2$ `hard' sources with a 0.2--2~keV count rate near the 
detection limit mentioned above. This additional information can be
used to distinguish TDE flares from e.g. AGN variability.  

To summarize, we adopt a detection limit of 20~counts inside the HPD
region during a 240~s exposure, i.e. $C_{lim}= 0.167$~cts~s$^{-1}$ in
the 0.2--2~keV energy band. We propose $SR=C_{0.4-1}/C_{1-2}=2$ as a
boundary between 'soft` and `hard' sources, which will allow one to
retain as many as possible true candidates while rejecting a
significant fraction of bogus ones. It might be necessary to
  slightly modify these flux and softness ratio criteria if the 
  background level measured during the eRASS proves 
  significantly different from the current estimates and/or typical
  TDE absorbing columns prove significantly different from the $N_H$
  value assumed above. 

\section*{Acknowledgements}
The research made use of grant NSh-5603.2012.2 from President of Russian
Federation, RFBR grants 13-02-12250-ofi-m and 13-02-01365, programmes P-21 and OFN-17 of the Russian Academy of Sciences and grant 87019 from Ministry of
Science and Education of Russian Federation. IK acknowledges the support of the Dynasty Foundation.


\begin{thebibliography}{99}

\bibitem[\protect\citeauthoryear{Ayal, Livio, 
\& Piran}{2000}]{Ayal2000} Ayal S., Livio M., Piran T., 2000, ApJ, 545, 772 

\bibitem[\protect\citeauthoryear{Alexander}{2012}]{Alexander2012} 
Alexander T., 2012, EPJWC, 39, 5001 

\bibitem[\protect\citeauthoryear{Bower et al.}{2012}]{Bower2012} 
Bower G.~C., Metzger B.~D., Cenko S.~B., Silverman J.~M., Bloom J.~S., 
2012, arXiv, arXiv:1210.0020 

\bibitem[\protect\citeauthoryear{Brandt 
\& Hasinger}{2005}]{Brandt2005} Brandt W.~N., Hasinger G., 2005,
  ARA\&A, 43, 827  

\bibitem[\protect\citeauthoryear{Brockamp, Baumgardt, 
\& Kroupa}{2011}]{Brockamp2011} Brockamp M., Baumgardt H., Kroupa P.,
  2011, MNRAS, 418, 1308  

\bibitem[\protect\citeauthoryear{Burrows et 
al.}{2011}]{Burrows2011} Burrows D.~N., et al., 2011, Natur, 476, 
421 

\bibitem[\protect\citeauthoryear{Cappelluti et 
al.}{2009}]{Cappelluti2009} Cappelluti N., et al., 2009, A\&A, 495, L9 

\bibitem[\protect\citeauthoryear{Carrasco et 
al.}{1998}]{Carrasco1998} Carrasco L., Tovmassian H.~M., Stepanian 
J.~A., Chavushyan V.~H., Erastova L.~K., Vald{\'e}s J.~R., 1998, AJ, 115, 
1717 

\bibitem[\protect\citeauthoryear{Cenko et al.}{2012}]{Cenko2012} 
Cenko S.~B., et al., 2012, ApJ, 753, 77 

\bibitem[\protect\citeauthoryear{Churazov et 
al.}{1996}]{Churazov1996} Churazov E., Gilfanov M., Forman W., Jones 
C., 1996, ApJ, 471, 673 
\bibitem[Donley et al.(2002)]{Donley2002} Donley, J.~L., Brandt, 
W.~N., Eracleous, M., \& Boller, T.\ 2002, ApJ, 124, 1308 

\bibitem[Dorman 
\& Arnaud(2001)]{Dorman2001} Dorman, B., \& Arnaud, K.~A.\ 2001,
  Astronomical Data Analysis Software and Systems X, 238, 415  

\bibitem[Esquej et 
al.(2008)]{Esquej2008} Esquej, P., Saxton, R.~D., Komossa, S., et
  al.\ 2008, A\&A, 489, 543  
  
\bibitem[\protect\citeauthoryear{Evans 
\& Kochanek}{1989}]{Evans1989} Evans C.~R., Kochanek C.~S., 1989, ApJ, 346, L13 

\bibitem[\protect\citeauthoryear{Gezari et al.}{2009}]{Gezari2009} 
Gezari S., et al., 2009, ApJ, 698, 1367 

\bibitem[\protect\citeauthoryear{Giannios 
\& Metzger}{2011}]{Giannios2011} Giannios D., Metzger B.~D., 2011,
  MNRAS, 416, 2102  

\bibitem[\protect\citeauthoryear{Greene 
\& Ho}{2007}]{Greene2007} Greene J.~E., Ho L.~C., 2007, ApJ, 667, 131 

\bibitem[\protect\citeauthoryear{Guillochon 
\& Ramirez-Ruiz}{2013}]{Guillochon2013} Guillochon J., Ramirez-Ruiz E., 2013, ApJ, 767, 25 


\bibitem[\protect\citeauthoryear{Gurzadian 
\& Ozernoi}{1981}]{Gurzadian1981} Gurzadian V.~G., Ozernoi L.~M.,
  1981, A\&A, 95, 39  
\bibitem[\protect\citeauthoryear{Halpern, Gezari, 
\& Komossa}{2004}]{Halpern2004} Halpern J.~P., Gezari S., Komossa S., 2004, ApJ, 604, 572 
\bibitem[\protect\citeauthoryear{Hills}{1975}]{Hills1975} Hills 
J.~G., 1975, Natur, 254, 295 

\bibitem[Ho(2008)]{Ho2008} Ho, L.~C.\ 2008, ARA\&A, 46, 475 

\bibitem[Hogg(1999)]{Hogg1999} Hogg, D.~W.\ 1999, arXiv:astro-ph/9905116 

\bibitem[\protect\citeauthoryear{Hopkins, Richards, 
\& Hernquist}{2007}]{Hopkins2007} Hopkins P.~F., Richards G.~T.,
  Hernquist L., 2007, ApJ, 654, 731  


\bibitem[\protect\citeauthoryear{Kaiser et al.}{2010}]{Kaiser2010} 
Kaiser N., et al., 2010, SPIE, 7733,  
\bibitem[\protect\citeauthoryear{Kesden}{2012}]{Kesden2012} Kesden 
M., 2012, PhRvD, 86, 064026 

\bibitem[\protect\citeauthoryear{Khabibullin, Sazonov, 
\& Sunyaev}{2012}]{Khabibullin2012} Khabibullin I., Sazonov S., Sunyaev R., 2012, MNRAS, 426, 1819 

\bibitem[\protect\citeauthoryear{Komatsu et 
al.}{2011}]{Komatsu2011} Komatsu E., et al., 2011, ApJS, 192, 18 

\bibitem[Komossa(2002)]{Komossa2002} Komossa, S.\ 2002, Reviews in 
Modern Astronomy, 15, 27 
\bibitem[\protect\citeauthoryear{Komossa et 
al.}{2004}]{Komossa2004} Komossa S., Halpern J., Schartel N., 
Hasinger G., Santos-Lleo M., Predehl P., 2004, ApJ, 603, L17
\bibitem[\protect\citeauthoryear{Komossa}{2012}]{Komossa2012} 
Komossa S., 2012, AdAst, 2012

\bibitem[\protect\citeauthoryear{Kormendy 
\& Richstone}{1995}]{Kormendy1995} Kormendy J., Richstone D., 1995, ARA\&A, 33, 581 
\bibitem[\protect\citeauthoryear{Kormendy, Ho}{2001}]{Kormendy2001} Kormendy J., Ho L., 2001

\bibitem[\protect\citeauthoryear{Krolik 
\& Piran}{2012}]{Krolik2012} Krolik J.~H., Piran T., 2012, ApJ, 749, 92 

\bibitem[\protect\citeauthoryear{Laguna et al.}{1993}]{Laguna1993} 
Laguna P., Miller W.~A., Zurek W.~H., Davies M.~B., 1993, ApJ, 410, L83 

\bibitem[\protect\citeauthoryear{Lei 
\& Zhang}{2011}]{Lei2011} Lei W.-H., Zhang B., 2011, ApJ, 740, L27 

\bibitem[\protect\citeauthoryear{Levan et al.}{2011}]{Levan2011} 
Levan A.~J., et al., 2011, Sci, 333, 199

\bibitem[\protect\citeauthoryear{Lidskii 
\& Ozernoi}{1979}]{Lidskii1979} Lidskii V.~V., Ozernoi L.~M., 1979, SvAL, 5, 16 

\bibitem[\protect\citeauthoryear{Lin et al.}{2011}]{Lin2011} 
Lin D., Carrasco E.~R., Grupe D., Webb N.~A., Barret D., Farrell S.~A., 
2011, ApJ, 738, 52

\bibitem[Lodato et al.(2009)]{Lodato2009} Lodato, G., King, A.~R., 
\& Pringle, J.~E.\ 2009, MNRAS, 392, 332 

\bibitem[\protect\citeauthoryear{Lumb et 
al.}{2002}]{Lumb2002} Lumb D.~H., Warwick R.~S., Page M., De Luca A.,
  2002, A\& A, 389, 93 

\bibitem[\protect\citeauthoryear{MacLeod, Guillochon, 
\& Ramirez-Ruiz}{2012}]{MacLeod2012} MacLeod M., Guillochon J., Ramirez-Ruiz E., 2012, ApJ, 757, 134 

\bibitem[\protect\citeauthoryear{Maksym, Ulmer, 
\& Eracleous}{2010}]{Maksym2010} Maksym W.~P., Ulmer M.~P., Eracleous
  M., 2010, ApJ, 722, 1035  

\bibitem[\protect\citeauthoryear{Massaro et 
al.}{2011}]{Massaro2011} Massaro F., D'Abrusco R., Ajello M., 
Grindlay J.~E., Smith H.~A., 2011, ApJ, 740, L48 

\bibitem[\protect\citeauthoryear{Merloni et 
al.}{2012}]{Merloni2012} Merloni A., et al., 2012, arXiv, 
arXiv:1209.3114 

\bibitem[\protect\citeauthoryear{Metzger, Giannios, 
\& Mimica}{2012}]{Metzger2012} Metzger B.~D., Giannios D., Mimica P.,
  2012, MNRAS, 420, 3528  

\bibitem[\protect\citeauthoryear{Nikolajuk 
\& Walter}{2013}]{Nikolajuk2013} Nikolajuk M., Walter R., 2013, arXiv,
  arXiv:1304.0397  

\bibitem[\protect\citeauthoryear{Pavlinsky et al.}{2012}]{Pavlinsky2012}
  Pavlinsky M., et al., 2012, SPIE Proc., in press
  
\bibitem[Peebles(1993)]{Peebles1993} Peebles, P.~J.~E.\ 1993, 
Principles of Physical Cosmology by P.J.E.~Peebles.~Princeton University 
Press, 1993.~ISBN: 978-0-691-01933-8  

\bibitem[\protect\citeauthoryear{Phinney}{1989}]{Phinney1989} 
Phinney E.~S., 1989, IAUS, 136, 543 

\bibitem[Rees(1988)]{Rees1988} Rees, M.~J.\ 1988, Nature, 333, 523 

\bibitem[\protect\citeauthoryear{Saxton et 
al.}{2012}]{Saxton2012} Saxton R.~D., Read A.~M., Esquej P., Komossa S., Dougherty S., Rodriguez-Pascual P., Barrado D., 2012, A\&A, 541, A106 

\bibitem[\protect\citeauthoryear{SDSS-III Collaboration et 
al.}{2012}]{SDSS2012} SDSS-III Collaboration, et al., 2012, 
arXiv, arXiv:1207.7137 

\bibitem[\protect\citeauthoryear{Sembay 
\& West}{1993}]{Sembay1993} Sembay S., West R.~G., 1993, MNRAS, 262, 141 

\bibitem[Shakura 
\& Sunyaev(1973)]{SS1973} Shakura, N.~I., \& Sunyaev, R.~A.\ 1973,
  A\&A, 24, 337  

\bibitem[\protect\citeauthoryear{Snowden et 
al.}{1997}]{Snowden1997} Snowden S.~L., et al., 1997, ApJ, 485, 125 

\bibitem[\protect\citeauthoryear{Stone, Sari, 
\& Loeb}{2012}]{Stone2012} Stone N., Sari R., Loeb A., 2012, arXiv, arXiv:1210.3374 


\bibitem[Strubbe 
\& Quataert(2009)]{Strubbe2009} Strubbe, L.~E., \& Quataert, E.\ 2009,
  MNRAS, 400, 2070  

\bibitem[Ulmer(1999)]{Ulmer1999} Ulmer, A.\ 1999, ApJ, 514, 180 


\bibitem[\protect\citeauthoryear{van Velzen et 
al.}{2011b}]{vanVelzen2011A} van Velzen S., et al., 2011, ApJ, 741, 73 

\bibitem[\protect\citeauthoryear{van Velzen, K{\"o}rding, 
\& Falcke}{2011}]{vanVelzen2011B} van Velzen S., K{\"o}rding E.,
  Falcke H., 2011, MNRAS, 417, L51  

\bibitem[\protect\citeauthoryear{van Velzen et 
al.}{2012}]{vanVelzen2012} van Velzen S., Frail D.~A., Koerding E., 
Falcke H., 2012, arXiv, arXiv:1210.0022 

\bibitem[\protect\citeauthoryear{Vaughan, Edelson, 
\& Warwick}{2004}]{Vaughan2004} Vaughan S., Edelson R., Warwick R.~S.,
  2004, MNRAS, 349, L1  

\bibitem[\protect\citeauthoryear{Veilleux 
\& Osterbrock}{1987}]{Veilleux1987} Veilleux S., Osterbrock D.~E.,
  1987, ApJS, 63, 295 

\bibitem[Wall et al.(2003)]{Wall2003} Wall, J.~V., Jenkins, 
C.~R., Ellis, R., et al.\ 2003, Practical statistics for astronomers, by 
J.V.~Wall and C.R.~Jenkins.~Cambridge observing handbooks for research 
astronomers, vol.~3.~Cambridge, UK: Cambridge University Press, 2003 

\bibitem[\protect\citeauthoryear{Wang 
\& Merritt}{2004}]{Wang2004} Wang J., Merritt D., 2004, ApJ, 600, 149 

\bibitem[\protect\citeauthoryear{Warwick, Saxton, 
\& Read}{2012}]{Warwick2012} Warwick R.~S., Saxton R.~D., Read A.~M., 2012, A\&A, 548, A99

\bibitem[\protect\citeauthoryear{Wright et al.}{2010}]{Wright2010} 
Wright E.~L., et al., 2010, AJ, 140, 1868 




\end{thebibliography}
\end{document}